\documentclass[conference]{IEEEtran}

\usepackage{cite}
\usepackage{amsmath,amssymb,amsfonts}
\usepackage{graphicx}
\usepackage{textcomp}
\usepackage{xcolor}
\usepackage{array}
\usepackage{url}

\def\BibTeX{{\rm B\kern-.05em{\sc i\kern-.025em b}\kern-.08em
    T\kern-.1667em\lower.7ex\hbox{E}\kern-.125emX}}

\begin{document}

\title{Auditable Graph-Guided Root Cause Analysis for Kubernetes Incidents}

\author{
\IEEEauthorblockN{Anastasiia Kuvshinova}
\IEEEauthorblockA{HSE University \\
Moscow, Russia \\
anastasiiakuvshinova02@gmail.com}
\and
\IEEEauthorblockN{Seungmin Jin\textsuperscript{*}}
\IEEEauthorblockA{HSE University \\
Moscow, Russia \\
sedzhin@hse.ru}
}

\maketitle
\begingroup
\renewcommand{\thefootnote}{*}
\footnotetext{Corresponding author.}
\endgroup
\begingroup
\renewcommand{\thefootnote}{}
\footnotetext[0]{A version of this work was accepted to the GNN-LLM-SYN workshop at ICDM 2026 but was not published in the workshop proceedings.}
\endgroup

\begin{abstract}
Kubernetes incidents are diagnosed reliably only when a root-cause system's reported gains come from incident evidence rather than scenario-specific shortcuts. We present Graph Traversal Agent, a graph-guided RCA agent that follows a pattern recommended in recent LLM-for-AIOps literature: combine LLM reasoning with specialized tools instead of asking the model to diagnose from text alone \cite{zhang2026aiops}. The model reasons over a typed evidence graph; deterministic graph and tool operations collect evidence, bound the search, and check proposed verdicts. We map operational constraints (read-only evidence collection, propagation-aware diagnosis, bounded execution, and independently validated verdicts) to a typed incident graph, a LangGraph traversal state machine, and a separate validation stage. On ITBench snapshots scored by one fixed \texttt{qwen-plus} judge, the audited system raises root-cause-entity F1 over an earlier iteration of the same system from $0.6087$ to $0.9130$ on a 23-scenario common subset. A prompt-level ablation then separates prompt-tuned gains from gains that survive once scenario-specific hints are removed (the stripped-prompt configuration retains $0.6958$ F1 on a 19-scenario subset); the surviving gain concentrates on ChaosMesh scenarios whose ground-truth root cause is the injected fault object already present in the evidence graph, so we report it as benchmark-coupled rather than broad cross-cluster RCA evidence. Lightweight checks (same-judge comparison, prompt-level ablation, cascade-source checking, and a telemetry no-leak test) mark claims as supported, pending, or out of scope. We scope the work to ITBench OpenTelemetry-demo snapshots. Live-cluster trials served as an engineering stress test, but alert state and trace availability did not stay stable enough for controlled scoring, so we make no production-readiness or mean-time-to-repair claim.
\end{abstract}



\section{Introduction}

Modern cloud applications fail through interactions among services, workloads, configuration objects, observability pipelines, and fault injectors. A root cause analysis (RCA) agent that only reports a candidate entity can look accurate while exploiting benchmark-specific identifiers, a deterministic fallback rule that supplies an answer when the agent itself reaches none, or a scoring judge more lenient than the one the baseline was given. This paper studies RCA as an applied systems problem in which performance, architecture decisions, and auditability must be reported together.

The system under study, Graph Traversal Agent, represents an incident snapshot as a typed graph $G_s=(V_s,E_s)$ whose nodes include Kubernetes resources, services, alerts, trace-derived entities, and fault-injection objects. The agent traverses this graph on a LangGraph state machine, invokes read-only evidence-gathering tools on candidate nodes, validates each root-cause classification against the collected evidence, and emits a classified candidate set with an explanation chain. Recent LLM-for-AIOps work argues for tool-augmented and hybrid designs: tools provide current operational data, while LLMs handle interpretation across noisy evidence sources \cite{zhang2026aiops}. Graph Traversal Agent follows that split. The central question is then which part of a benchmark gain survives an audit against shortcut behavior and benchmark coupling, that is, gains that come from the benchmark's specific scenarios rather than from general diagnostic ability.

This paper makes three applied contributions.
\begin{itemize}
\item It maps production RCA requirements to concrete architecture choices: typed evidence ingestion, graph-guided traversal, read-only tool use, deterministic routing, checkpointing, and a separate validation stage.
\item It describes a graph-guided RCA pipeline for Kubernetes incidents, implemented as a LangGraph state machine, including graph construction, triage, candidate traversal, investigation, validation, and chain construction.
\item It introduces a set of audit checks that co-report same-judge snapshot metrics, prompt-level ablations, cascade-source checks, telemetry no-leak tests, and positive and negative case studies.
\end{itemize}

The remaining sections proceed as follows. Section~\ref{sec:related} positions the work against prior RCA systems. Section~\ref{sec:problem} defines the RCA task. Section~\ref{sec:system} describes the system. Section~\ref{sec:audit} gives the audit checks and the shortcuts they detect. Section~\ref{sec:snapshot} reports snapshot evidence. Section~\ref{sec:live} reports the live-validation status. Section~\ref{sec:failures} presents the case studies and failure analysis. Section~\ref{sec:limitations} states the remaining scope limits.

\section{Related Work}
\label{sec:related}

RCA papers do not all solve the same task. Fang et al. organize cloud RCA around goal structure rather than telemetry type, distinguishing point localization from richer outputs such as propagation structure and actionable explanations \cite{fang2025goal}. ITBench similarly exposes multiple SRE evaluation dimensions, including root-cause entity matching, propagation chain quality, fault localization, and reasoning quality \cite{jha2025itbench}. This paper follows that separation: root-cause accuracy is not treated as a substitute for propagation evidence, and every reported gain must be tied to the judged dimension and scenario subset that produced it.

Kubernetes incidents add a failure surface that is not captured by service-level metrics alone. Barletta et al. show that Kubernetes failures propagate across orchestrator and client-visible layers, which motivates collecting evidence from resource state, events, metrics, and traces rather than from a single telemetry stream \cite{barletta2024mutiny}. SynergyRCA also uses graph structure for Kubernetes RCA by combining a StateGraph and a MetaGraph to constrain LLM reasoning over valid Kubernetes entities \cite{xiang2025synergyrca}. Our system is closer to this Kubernetes-specific line than to generic log-only or metric-only RCA, but the current evidence still remains bounded to the OpenTelemetry-demo and ChaosMesh setting.

LLM-based RCA systems have explored several ways to make language models operationally useful. ReAct introduced interleaved reasoning and acting as a general tool-use pattern \cite{yao2023react}. RCACopilot applies LLMs to cloud incident RCA through alert handling, evidence aggregation, root-cause category prediction, and narrative generation \cite{chen2023rcacopilot}; RCAgent studies autonomous tool use for cloud RCA with context management and trajectory stabilization \cite{wang2024rcagent}; COCA uses code knowledge to supplement incomplete issue reports for distributed-system failures \cite{li2025coca}. These systems motivate tool-grounded diagnosis, but their reported metrics are not directly comparable to our ITBench snapshot numbers unless the same benchmark, judge, and scenario subset are used.

For this reason, the evaluation prioritizes controlled evidence over cross-paper score comparison. Quantitative claims are reserved for outputs that share the benchmark subset, judge model, output schema, and run artifacts; systems whose public results do not share these controls are discussed as design context.

The AIOps survey by Zhang et al. provides the design rationale for this paper's LLM-plus-tools structure \cite{zhang2026aiops}. It separates prompt-only methods from knowledge-based approaches such as retrieval-augmented generation and tool-augmented generation. For AIOps, tool-augmented methods let an LLM query monitoring systems, diagnostic scripts, and debugging tools. The same survey also argues that practical systems should integrate LLMs with smaller models, rule-based components, legacy systems, and specialized tools so routine processing stays outside the LLM and the model handles interpretation. Graph Traversal Agent uses that division: graph construction, queue management, tool execution, and validation checks are deterministic, while the LLM interprets node evidence and proposes hypotheses.

Graph-guided RCA is the closest methodological neighbor. PRAXIS constrains an LLM-driven workflow with service-dependency and program-dependence graphs for code-related cloud incidents \cite{cui2026praxis}, while SynergyRCA constrains Kubernetes diagnosis through graph database retrieval and expert prompts \cite{xiang2025synergyrca}. The present paper uses a typed incident graph and bounded traversal for Kubernetes evidence. Its added emphasis is claim auditing: same-judge comparisons, prompt-level ablations, cascade-source checks, telemetry no-leak tests, and explicit failure reports.

Multi-agent RCA work shows both opportunity and risk. Flow-of-Action uses SOP-guided multi-agent coordination for RCA \cite{pei2025flow}, mABC studies blockchain-inspired multi-agent voting \cite{zhang2024mabc}, and STRATUS organizes specialized reliability agents in a state machine for autonomous SRE tasks \cite{chen2026stratus}. At the same time, MAST identifies recurring multi-agent failure modes \cite{cemri2025mast}, and cloud-RCA-specific failure analysis finds persistent incomplete exploration and hallucinated evidence interpretation across agent runs \cite{kim2026cloudrcafailures}. These findings motivate our conservative framing: verification and audit are reported as mechanisms for exposing failure modes, not as evidence that a multi-agent architecture is inherently superior.

\section{Problem Definition}
\label{sec:problem}

\subsection{Incident RCA Task}

For each incident scenario $s$, the agent observes an evidence bundle $X_s$ and a graph $G_s=(V_s,E_s)$. The output is a set of classified candidate root-cause entities and a propagation chain. A prediction is useful only if the named entity is supported by tool-observed evidence and if the proposed chain is compatible with the incident propagation evidence.

The evaluation uses four ITBench judge dimensions: root-cause entity identification, propagation chain, fault localization, and root-cause reasoning \cite{jha2025itbench}. Each reported table states its judge model, scenario subset, and run identifiers.

The defensible scope of this study is RCA on incidents injected by ChaosMesh (a Kubernetes fault-injection tool) into the OpenTelemetry Demo application (a standard reference microservice deployment); it is not RCA that generalizes across arbitrary clusters, application stacks, and fault injectors. The shortcuts that would invalidate a general-ability claim, and the checks that detect them, are defined in Section~\ref{sec:audit}.

\section{System Overview}
\label{sec:system}

The pipeline is implemented as a LangGraph state machine \cite{langgraph}: graph construction, triage, breadth-first traversal, per-node investigation, validation, optional strategic re-ranking, and chain construction are nodes in a directed state graph, with an SQLite checkpointer providing per-step persistence and resume. The agents do not converse freely; all interaction is mediated through the shared state object.

\subsection{Evidence Graph}

The graph builder combines Kubernetes resources, service relationships, alert labels, trace-derived dependencies, logs, events, and fault-injection objects into a typed directed multigraph. Edges use six typed relations, each annotated with its evidence source: \texttt{owns}, \texttt{hosts}, \texttt{selects}, \texttt{mounts}, \texttt{fronts}, and \texttt{calls} (Table~\ref{tab:req-arch}). The graph gives the agent a bounded search surface instead of a free-form text context.

\subsection{Design Constraints}

Table~\ref{tab:req-arch} lists the operational constraints that shaped the prototype. Each row states the constraint, the implementation choice, and the artifact a reviewer can inspect.

\begin{table*}[t]
\caption{Operational constraints mapped to implementation choices.}
\label{tab:req-arch}
\centering
\setlength{\tabcolsep}{2pt}
\small
\begin{tabular}{p{0.15\textwidth}p{0.28\textwidth}p{0.34\textwidth}p{0.14\textwidth}}
\hline
Constraint & Operational need & Implementation choice & Evidence \\
\hline
Read-only evidence & Inspect cluster state without changing it. & Tools use Kubernetes, Prometheus, Jaeger, event, and log queries; Secret access is blocked. & Tool allow-list \\
Typed graph & Preserve service, workload, config, node, and trace relationships. & A NetworkX multigraph stores \texttt{owns}, \texttt{hosts}, \texttt{selects}, \texttt{mounts}, \texttt{fronts}, and \texttt{calls} edges with source labels. & Graph outputs \\
Bounded search & Avoid sending the LLM an unbounded evidence dump. & LangGraph state stores alerts, queue, current node, evidence log, classifications, validation state, and chain accumulator. & Checkpoints \\
Validation control & Expose an investigator-only configuration without using it as a scored claim here. & The runner exposes \texttt{--no-validation}; the reported tables keep validation enabled. & CLI option \\
Verdict checks & Prevent a single investigator classification from becoming final without review. & ValidationAgent checks temporal order, evidence grounding, ConfigMap symptoms, and cross-modal consistency, then accepts, retries, or redirects. & Validation traces \\
Judge output & Produce structured RCA output, not only prose. & Chain construction writes ITBench-compatible root-cause candidates, classifications, alert explanations, and propagation hops. & \texttt{agent\_output.json} \\
Replay & Make snapshot runs inspectable after the fact. & Snapshot replay uses archived evidence; each run records output directories and model-call telemetry. & Run directories \\
\hline
\end{tabular}
\end{table*}

\subsection{Triage and Traversal}

A triage stage uses active AlertManager events to reorder the breadth-first queue so likely candidates go first. For alert patterns with an unambiguous Kubernetes target, SOP rules can pin that target to the front of the queue; otherwise a single pre-traversal model call proposes the initial focus. Triage narrows the order of investigation. It does not by itself supply the scored answer reported in the tables. The traversal layer then visits candidate nodes under a bounded breadth-first budget. For each candidate, the investigator agent issues a ReAct cycle over a fixed read-only tool allow-list (Kubernetes object lookup, events, logs, Prometheus queries, and Jaeger queries) and classifies the node as a \emph{root cause}, \emph{intermediate}, \emph{symptom}, or \emph{unrelated} entity, recording a fault signal and a categorical confidence when available.

\subsection{Validation and Chain Construction}

A separate validation stage gates each root-cause classification. It first applies deterministic checks (temporal precedence: cause evidence must not post-date the first alert; evidence grounding: the cited fault signal must resolve to a concrete observed record; cross-modal consistency: an active metric alert must be corroborated by trace behaviour) and rejects classifications that fail them. Ambiguous cases can enter a director-executor validation path: a validation director proposes one read-only tool query, an executor runs that query, and the validator accepts, asks for a retry, or redirects the search. The chain builder then emits the classified candidate set and a propagation chain for the external judge. The audit distinguishes investigator-produced positions from deterministic fallback or cascade-promotion positions.

\begin{figure*}
\centering
\resizebox{\textwidth}{!}{\includegraphics{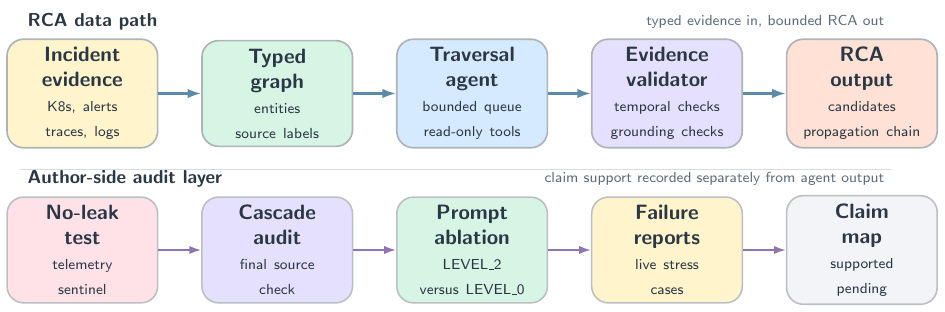}}
\caption{System overview. The RCA data path produces classified root-cause candidates and a propagation chain from typed incident evidence. The author-side audit package records which claims are supported by run outputs, judge files, ablations, tests, or failure reports, and which claims remain out of scope.}
\label{fig:system}
\end{figure*}

\section{Audit Checks}
\label{sec:audit}

The audit checks are author-side reproducibility and claim-governance checks, not an autonomous auditor inside the RCA agent. They report performance only when the supporting evidence path is explicit. Table~\ref{tab:evidence-map} maps each planned claim to its required evidence unit.

\begin{table}
\caption{Claim-evidence map for the paper.}
\label{tab:evidence-map}
\centering
\setlength{\tabcolsep}{2pt}
\begin{tabular}{p{0.27\columnwidth}p{0.33\columnwidth}p{0.30\columnwidth}}
\hline
Claim & Evidence unit & Status \\
\hline
Snapshot accuracy improves under the same judge & ITBench snapshot sweep, common scenario subset, same judge output & Available \\
ChaosMesh handling survives prompt-level ablation & LEVEL\_0 versus LEVEL\_2 ablation on common chaos subset & Available for current scope \\
Reported gains are not cascade shortcuts & Cascade-source fraction per run & Check implemented; final report artifact pending \\
Telemetry cannot leak into prompts & Unit tests with sentinel strings and concurrent thread isolation & Available \\
Live validation supports deployed operation & Five live ChaosMesh stress-test scenarios & Not claimed: trial isolation and trace availability were unstable \\
Case-study analysis improves system understanding & Scenario 9 positive snapshot trace plus Scenario 16 live failure reports & Available for current scope \\
\hline
\end{tabular}
\end{table}

\subsection{Shortcuts the Checks Detect}

The checks treat the following as invalid evidence of general RCA ability:
\begin{itemize}
\item Scenario leakage through names, prompt hints, or benchmark-specific service identifiers.
\item Cascade promotion: a fallback rule that names a root cause by promoting a traversed graph node when the investigator agent produces no verdict, counted as if it were agent reasoning. We do not forbid this fallback; the cascade-source check below measures its share and credits a gain only when that share is low.
\item Judge variance from comparing outputs scored by different judge models.
\item Telemetry leakage, where run metadata is appended to model prompts.
\item Fault-injector coupling, where ChaosMesh-specific constants are mistaken for general fault-injection competence.
\end{itemize}

\subsection{Same-Judge Snapshot Comparison}

All snapshot comparisons must use a common scenario subset judged by the same judge model. Cross-judge comparisons are reported only as a source of variance, not as performance improvement.

\subsection{Prompt-Level Ablation}

The agent's prompt has graded levels of scenario-specific guidance. $\text{LEVEL\_2}$ is the default used for the headline results: it includes fault-class-specific instruction blocks (for example, hints for feature-flag and ChaosMesh faults). $\text{LEVEL\_0}$ strips these, leaving only generic node-classification rules, the read-only tool reference, and the response format. The ablation compares $\text{LEVEL\_2}$ with $\text{LEVEL\_0}$ on the same code iteration. It is not a perfect architecture-agnostic ablation because deterministic code paths still include ChaosMesh- and flagd-specific constants; the paper states this residual coupling before interpreting the ablation.

\subsection{Cascade-Source Check}

For each judged output, the audit computes how many candidate positions came from investigator-agent verdicts versus fallback mechanisms. Each position is classified as agent-committed (named as a root cause by an investigator verdict), diagnostic (the agent visited the entity but did not name it root cause), or cascade-promoted (filled by a fallback mechanism that bypasses the agent's verdict; see the shortcuts above). Under the intended gating rule, a performance claim is credited only when the cascade-promoted fraction is below a preset threshold. For the runs reported here, the per-run cascade-source report has not been generated, so the reported snapshot gains are not yet gated by this check; we present it as part of the audit checks, not a completed result (Table~\ref{tab:evidence-map}).

\subsection{Telemetry No-Leak Audit}

The telemetry wrapper writes model-call metadata to trial-local JSON lines files via thread-local context. Sentinel tests check that scenario, trial, and telemetry strings are not appended to the prompt payload. This audit supports instrumentation safety, not RCA accuracy.

\section{Snapshot Evaluation}
\label{sec:snapshot}

\subsection{Experimental Setup}

The snapshot benchmark uses ITBench incident snapshots and the unmodified external judge harness \cite{jha2025itbench}. The judge for every table reported here is \texttt{qwen-plus} and the agent backend is \texttt{qwen-flash}, both served through the DashScope API; the runtime can substitute an interchangeable hosted model when a provider quota limit is hit, and the model actually used for each scored call is recorded in the per-call telemetry log. We report the exact code iteration, prompt level, backend model, judge model, completed scenarios, failed scenarios, timeout scenarios, and run directories.

The comparison uses three archived runs, referred to by role: a \emph{prior-iteration baseline}, the \emph{audited system} at its default prompt level ($\text{LEVEL\_2}$), and the audited system re-run at $\text{LEVEL\_0}$. Their internal development identifiers are iter-7, iter-3, and iter-10 respectively; these are sequential development tags, not chronological indices, and the prior-iteration baseline (iter-7) is the earlier system despite its higher tag.

\subsection{Same-Judge Results}

Table~\ref{tab:same-judge} reports a same-judge, same-prompt-level code-iteration comparison on the two-way common subset shared by the prior iteration and the audited system: S2, S4, S5, S7, S8, S9, S11, S12, S13, S14, S15, S16, S17, S18, S19, S23, S24, S25, S33, S35, S38, S81, and S105. A common subset means the scenarios for which every compared configuration produced a scored judge output. Every value is traceable to a judge output file. The \emph{prior iteration} column (iter-7) is an earlier iteration of our own system at the same prompt level and \texttt{qwen-plus} judge as the audited system (iter-3); the improvement reflects accumulated code and prompt refinements, not a single added component. This is an intra-system, same-judge delta, not the published ITBench leaderboard baseline (the GPT-OSS-120B reference agent at 0.164 root-cause-entity F1 under that benchmark's \texttt{gemini-2.5-pro} judge); cross-judge magnitudes are excluded.

\begin{table}
\caption{Same-judge snapshot comparison ($n{=}23$; \texttt{qwen-plus} judge; same prompt level). \emph{Prior iter.} (iter-7) is an earlier iteration of our own system, not the ITBench leaderboard baseline (see text).}
\label{tab:same-judge}
\centering
\setlength{\tabcolsep}{2pt}
\begin{tabular}{p{0.39\columnwidth}p{0.18\columnwidth}p{0.18\columnwidth}p{0.15\columnwidth}}
\hline
Metric & Prior iter. & Audited system & Delta \\
\hline
Root-cause entity F1 & 0.6087 & 0.9130 & +30.4 pp \\
Propagation chain & 0.3448 & 0.4028 & +5.8 pp \\
Fault localization & 0.4348 & 0.4783 & +4.4 pp \\
Root-cause reasoning & 0.6087 & 0.9130 & +30.4 pp \\
\hline
\end{tabular}
\end{table}

The interpretation of Table~\ref{tab:same-judge} is limited: it is an observed same-prompt-level code-iteration delta on a common subset, not evidence of architecture-agnostic generalization or of a single causal component. It is also not yet gated by the cascade-source check in Section~\ref{sec:audit}; the table should be read as a same-judge performance result whose provenance audit remains an explicit artifact requirement. The prompt-level ablation in Table~\ref{tab:level0} is the main scope-bounding table, because it separates prompt-tuned gains from gains that survive the prompt-level-zero setting.

\subsection{Prompt-Level Ablation}

Table~\ref{tab:level0} separates prompt-tuned gains from gains that survive the prompt-level-zero setting on the three-way common subset shared by the prior iteration, LEVEL\_2 run, and LEVEL\_0 run: S2, S4, S5, S7, S9, S11, S12, S13, S14, S15, S16, S17, S18, S23, S24, S33, S35, S81, and S105. Compared with Table~\ref{tab:same-judge}, scenarios S8, S19, S25, and S38 are excluded because no LEVEL\_0 scored output was available for them in the archived run.

\begin{table}
\caption{Prompt-level ablation ($n=19$; \texttt{qwen-plus} judge). \emph{Prior iter.} (iter-7) is an earlier iteration of our own system, not the ITBench leaderboard baseline (see text).}
\label{tab:level0}
\centering
\setlength{\tabcolsep}{2pt}
\begin{tabular}{p{0.35\columnwidth}p{0.16\columnwidth}p{0.16\columnwidth}p{0.16\columnwidth}}
\hline
Metric & Prior iter. & LEVEL\_2 & LEVEL\_0 \\
\hline
Root-cause entity F1 & 0.6842 & 0.9474 & 0.6958 \\
Propagation chain & 0.4092 & 0.4504 & 0.2881 \\
Fault localization & 0.5263 & 0.5263 & 0.3684 \\
Root-cause reasoning & 0.6842 & 0.9474 & 0.7368 \\
\hline
\end{tabular}
\end{table}

At LEVEL\_0, the net root-cause-entity F1 over the prior iteration is about $+1.2$ pp, so most of the aggregate gain is prompt-tuned, while the five ChaosMesh official-injector cases remain solved. Entity F1 also runs above fault localization ($0.95$ versus $0.53$ at LEVEL\_2), because the labeled root cause is often the benchmark-visible injector, which the system names without localizing the first impacted service; the entity score should be read alongside fault localization, as the judge intends \cite{jha2025itbench}. This supports a narrow benchmark-scope statement: official ChaosMesh-object selection in these OpenTelemetry-demo snapshots is retained after removing scenario-specific prompt blocks. It does not support a broader claim about arbitrary feature-flag systems, arbitrary fault injectors, or cross-cluster generalization.

\subsection{ChaosMesh Subset}

The chaos subset is the clearest retained result, with one validity caveat. In these ITBench snapshots the ground-truth root-cause entity is the injected ChaosMesh object itself, and that object is present as a node in the evidence graph the agent reads. The result measures selection of the correct official injector object in the graph. It does not establish recovery of a hidden production cause from symptoms alone. We therefore report it as benchmark-scope evidence for the official ChaosMesh-labelled task. Defining a case as \emph{solved} when its root-cause-entity F1 equals $1.0$, the common five-scenario chaos pool has zero of five solved by the prior-iteration baseline, five of five solved by the audited system at $\text{LEVEL\_2}$, and five of five solved at $\text{LEVEL\_0}$. A production-realistic variant (injector object hidden, target re-defined as the affected service one hop downstream) is identified as future work (Section~\ref{sec:limitations}).

\section{Live-Validation Status}
\label{sec:live}

We ran a live-cluster stress test with five ChaosMesh scenarios on a Kubernetes cluster, including preflight checks, fault injection, cleanup, and post-trial evidence capture. The cluster did not provide a stable scoring environment: alerts sometimes carried over between trials, and trace data was intermittently unavailable. We therefore do not report a live success rate. All performance claims rest on the snapshot evaluation (Section~\ref{sec:snapshot}); the live runs are used as failure analysis (Section~\ref{sec:failures}) because they exposed observability and trial-isolation failure modes. Live trial logs are not part of the reproducible artifact; the reproducible results are the snapshot judge outputs.

\begin{table}
\caption{Live evidence status.}
\label{tab:live-status}
\centering
\setlength{\tabcolsep}{2pt}
\begin{tabular}{p{0.31\columnwidth}p{0.29\columnwidth}p{0.30\columnwidth}}
\hline
Evidence & Current status & Allowed claim \\
\hline
Five ITBench live ChaosMesh scenarios & Stress test; trial isolation and trace availability were unstable & None: no live success claimed \\
Scenario 16 live trials & Failure-mode evidence & Failure modes and repair targets \\
\hline
\end{tabular}
\end{table}

\section{Case Studies and Failure Analysis}
\label{sec:failures}

We include case studies because aggregate scores hide why the system succeeds or fails. We report one positive snapshot case where the graph design is necessary for success, and one negative case (Scenario 16, which the system solves on the snapshot but whose live trial failed) where the trial analysis localizes the failure.

\subsection{Scenario 9: Feature-Flag ConfigMap Fault}

Scenario 9 injects a feature-flag fault by changing the \texttt{loadGeneratorFloodHomepage} flag in the \texttt{flagd-config} ConfigMap from the off variant to the on variant. In the archived iter-3 snapshot run, the system's \texttt{agent\_output.json} names \texttt{otel-demo/ConfigMap/flagd-config} as the contributing root-cause entity, cites \texttt{loadGeneratorFloodHomepage=on} as the fault signal, and emits a propagation path from \texttt{flagd-config} to \texttt{flagd} and then to \texttt{frontend-proxy}. The judge scores this run with root-cause entity F1 $=1.0$, root-cause reasoning $=1.0$, propagation-chain score $=0.8$, and fault localization $=1.0$.

This case is positive evidence for the architecture, not only for prompt wording. The fault is not a service metric by itself; it is a configuration object that only becomes diagnosable if ConfigMaps are first-class graph nodes and if the investigator is forced to inspect mounted configuration data rather than only downstream service alerts.

\subsection{Scenario 16: Environment Variable Fault}

Scenario 16 is one of the live trials referenced in Section~\ref{sec:live}; it illustrates why the live-cluster runs were not used as a scored benchmark. Its ground truth root cause is \texttt{Deployment/shipping}, with \texttt{QUOTE\_ADDR} set to \texttt{quote:0000}. In one live trial, traversal reached the shipping deployment and called \texttt{k8s\_get}, but the relevant environment variable was truncated from the observation window and the validator rejected empty onset evidence. In a later trial, an orphaned \texttt{kubectl} port-forward bound to a terminated Jaeger pod silently stopped returning trace data, the investigator misread a valid Kubernetes service address as a misconfiguration, and the validator accepted a false \texttt{checkout} root cause while the correct node (\texttt{shipping}) was still unvisited in the traversal queue. The live trial logs are not part of the reproducible artifact.

\begin{table}
\caption{Scenario 16 failure decomposition.}
\label{tab:s16}
\centering
\setlength{\tabcolsep}{2pt}
\begin{tabular}{p{0.22\columnwidth}p{0.38\columnwidth}p{0.30\columnwidth}}
\hline
Layer & Failure & Evidence \\
\hline
Observation & Tool result truncated before \texttt{QUOTE\_ADDR} & Live trial 1 \\
Validation & Empty onset evidence rejected for env-var fault & Live trial 1 \\
Operations & Jaeger port-forward bound to dead pod & Live trial 2 \\
Investigation & Valid service DNS misread as env-var fault & Live trial 2 \\
Traversal & Alert-bearing nodes delayed behind mount nodes & Live trial 2 \\
\hline
\end{tabular}
\end{table}

The correct interpretation is negative but useful: these trials do not show live success, but they do show that the recorded tool calls, validation decisions, and trial reports can localize the source of wrong root causes across system layers. Together, Scenario~9 and Scenario~16 show both sides of the applied claim: the graph can expose configuration-layer root causes that service-only traversal misses, while the author-side audit checks prevent us from hiding live failures when evidence collection or validation breaks.

\section{Limitations and Future Work}
\label{sec:limitations}

The evidence has six limits.
\begin{itemize}
\item The snapshot benchmark is ITBench-specific and uses its snapshot schema.
\item The best retained gain is ChaosMesh-specific; non-ChaosMesh injectors are not demonstrated.
\item Feature-flag handling includes flagd-specific assumptions and does not generalize to arbitrary feature-flag systems.
\item The live five-scenario ChaosMesh stress test did not provide a stable scoring environment because alert state and trace availability varied between trials, so live reliability and operational benefit are not claimed.
\item The reported snapshot tables are logged single-run sweeps on fixed scenario subsets, not multi-trial estimates; the current paper therefore does not report means, confidence intervals, or variance across repeated LLM runs. ITBench defines Pass@$k$ and Majority@$k$ consistency metrics for repeated-trial evaluation \cite{jha2025itbench}; such estimates are future work here.
\item No production deployment evidence currently supports measured mean-time-to-repair reduction, production impact, or cross-cluster generalization. Any runtime or economic interpretation should therefore be read as prototype-stage projection, not observed production MTTR.
\end{itemize}

These limitations bound the paper's scope: the applied contribution is an auditable RCA system whose supported claims are separated from benchmark-aware behavior.

One future-work direction follows directly from the ChaosMesh-subset result. In the ITBench snapshots, the ground-truth root cause for chaos scenarios is the injected ChaosMesh object itself, which is present in the evidence graph. In a production cluster no such injector object exists; the same fault manifests only as downstream symptoms. A more production-realistic task therefore removes the injector objects from the evidence and re-defines the target as the affected service one hop downstream, diagnosed from propagation evidence. As a preliminary test of this variant, we re-labelled the target as the first affected service and made propagation-chain construction structural: a caller-direction cascade from the affected service to the ingress, rather than a forward fan-out through shared hubs. Under this author-defined ground truth, a single-trial run over the twelve ChaosMesh scenarios localized the affected service in eight scenarios, with a mean propagation-chain score of $0.755$ over localized cases under the same \texttt{qwen-plus} judge; five of twelve reached perfect root-cause-entity F1. The four remaining cases are network-partition and JVM faults whose cascade runs to a severed downstream callee the chain builder does not yet emit. Because this grades against a relabelled target rather than the external ITBench standard, and uses a modified chain builder, we report it only as exploratory evidence that this implementation sometimes localizes affected services under an author-defined target without visible injector objects, not as a headline result. The relabelled scenarios, regeneration scripts, and judged evidence are provided as supplementary material. The same injector-invisible protocol can be applied next to the feature-flag scenarios, whose ground-truth root cause is the \texttt{flagd-config} ConfigMap rather than an affected service.

\section{Reproducibility Checklist}
\label{sec:reproducibility}

The reproducibility package for this paper is defined by the artifact classes in Table~\ref{tab:artifact-manifest}. The artifact repository URL is \url{https://github.com/akuvshinova-ds/graph-traversal-agent-artifact}. The prior-iteration columns in Tables~\ref{tab:same-judge} and~\ref{tab:level0} are reproduced from the iter-7 evaluation output; that output is treated as a required artifact alongside the audited iter-3 and iter-10 outputs. Snapshot judge outputs and snapshot run scripts are required for reproducing the reported tables. Live-run scripts are protocol artifacts only because the live-cluster stress test did not provide a stable scoring environment; the live trials appear only as the Scenario~16 failure analysis (Section~\ref{sec:failures}).

\begin{table}
\caption{Artifact manifest for reported claims.}
\label{tab:artifact-manifest}
\centering
\setlength{\tabcolsep}{2pt}
\begin{tabular}{p{0.24\columnwidth}p{0.38\columnwidth}p{0.28\columnwidth}}
\hline
Claim class & Required artifact & Submission status \\
\hline
Same-judge snapshot tables & iter-7, iter-3, and iter-10 judge outputs for the stated common subsets & Required supplement \\
Prompt-level ablation & Prompt level, model, scenario list, and judge outputs for LEVEL\_2 and LEVEL\_0 & Required supplement \\
Telemetry no-leak & Sentinel unit tests and telemetry wrapper source & Required supplement \\
Cascade-source audit & Audit script plus per-run cascade-source report & Script included; report pending \\
Injector-invisible exploratory run & Relabelled targets, regeneration scripts, and judged outputs for the twelve ChaosMesh scenarios & Exploratory supplement \\
Live evidence & Injection protocol and Scenario 16 failure reports & Failure analysis only \\
\hline
\end{tabular}
\end{table}

The reproducibility checklist is completed under the following interpretation. The paper includes the model and algorithm description, task assumptions, dataset source, scenario subsets, evaluation metrics, and compute/runtime notes. The tables report exact logged runs rather than mean-and-variance estimates; this is marked as a limitation rather than hidden. Public artifacts exclude private credentials, local paths, cloud project identifiers, and unreleased service metadata.

\section{Conclusion}

Graph-guided RCA can improve Kubernetes incident diagnosis only when evaluation shows how the answer was produced. The audited system improves ITBench snapshot scores under a fixed judge, but the prompt-level ablation shows most of that gain is prompt-tuned. The retained result is benchmark-coupled ChaosMesh evidence, not a broad cross-cluster RCA result. The applied result is the system plus the checks that keep its accuracy claims bounded.

\section*{Acknowledgment}
The authors thank Hoon Jo for operational feedback on Kubernetes reliability engineering and deployment constraints.

This research was supported in part through computational resources of HPC facilities at HSE University.
The article was prepared within the framework of the HSE University Basic Research Program.

\bibliographystyle{IEEEtran}
\bibliography{references}

\end{document}